\documentclass[final,5p,times,twocolumn]{elsarticle} 

\usepackage{lineno,hyperref}
\modulolinenumbers[200]

\begin{document}

\begin{frontmatter}

\title{High-light-yield calcium iodide (CaI$_2$) scintillator for astroparticle physics}

\author[address1]{Takashi Iida\corref{mycorrespondingauthor}}
\address[address1]{Faculty of Pure and Applied Sciences, University of Tsukuba, 1-1-1 Tennodai, Tsukuba, Ibaraki 305-8571, Japan}
\cortext[mycorrespondingauthor]{Corresponding author}
\ead{tiida@hep.px.tsukuba.ac.jp}

\author[address2]{Kei Kamada}
\author[address3]{Masao Yoshino}
\author[address3]{Kyoung Jin Kim}
\author[address4,address5]{Koichi Ichimura}
\author[address2,address3]{Akira Yoshikawa}

\address[address2]{New Industry Creation Hatchery Center, Tohoku University, 6-6-10 Aoba, Sendai, Miyagi 980-8579, Japan}
\address[address3]{Institute for Material Research, Tohoku University, 2-1-1, Sendai, Miyagi 980-8577, Japan}
\address[address4]{Kamioka Observatory, Institute for Cosmic Ray Research, University of Tokyo, 456, Higashi-Mozumi, Kamioka, Hida, Gifu 506-1205, Japan, Japan}
\address[address5]{Kavli Institute for the Physics and Mathematics of the Universe (WPI), the University of Tokyo, Kashiwa, Chiba, 277-8582, Japan}

\begin{abstract}
We are developing a high-light-yield calcium iodide (CaI$_2$) scintillator for astroparticle physics experiments. This paper reports our CaI$_2$ scintillator crystal's scintillation performance. It achieved high light emissions (2.7 times those of NaI(Tl)) at a wavelength that was well within the photomultiplier's sensitivity region. We also performed a pulse shape discrimination study using alpha and gamma-ray sources, confirming after a brief analysis that CaI$_2$ has excellent pulse shape discrimination potential.

\end{abstract}


\end{frontmatter}

\linenumbers

\section{Introduction}

Underground astroparticle physics experiments, such as searches for dark matter or neutrino-less double beta decay, require either a low energy threshold or a high energy resolution. Due to their high light yields, inorganic scintillators are being widely adopted for use in astroparticle detectors.

There is abundant indirect evidence that strongly implies the existence of non-luminous dark matter \cite{Zwicky}. Based on cosmological observations, it is estimated that only 5\% of the universe is made of ordinary matter. However, despite 27\% of the universe being composed of dark matter, we still know little of its nature. Weakly Interacting Massive Particles (WIMPs) are one of the most promising candidates for particulate dark matter \cite{WIMP}. The DAMA/LIBRA experiment has observed an annual event rate modulation with high statistical significance using NaI(Tl) scintillator crystals  \cite{DAMA}, and has claimed to have discovered dark matter. That said, several other highly sensitive experiments have ruled out the DAMA result \cite{XENON1T}. These results should be investigated for a range of target nuclei. In order to observe an annual signal modulation caused by WIMP-induced nuclear recoil, it is very important to lower the energy threshold.

Searches for neutrino-less double beta decay (0$\nu\beta\beta$) are considered to be the most practical way of determining whether neutrinos are of Majorana or Dirac type and of discovering lepton number violation. If neutrinos are of Majorana type, i.e., neutrinos and anti-neutrinos are identical, then the leptogenesis scenario gives a good explanation of the matter--antimatter asymmetry in the present universe \cite{Leptogenesis}. To distinguish the mono-energetic electron pairs produced by 0$\nu\beta\beta$ from the continuous-energy electron pairs produced by two-neutrino double beta decay (2$\nu\beta\beta$) in 0$\nu\beta\beta$ experiments, it is important to have high energy resolution. The most stringent lower limit so far on the half-life of neutrino-less double beta decay is T$_{1/2} >$ 10$^{26}$ years (90\% C.L.), given by the KamLAND-Zen experiment \cite{KamLAND}. The CANDLES experiment is currently searching for $^{48}$Ca double beta decay using undoped CaF$_2$ scintillators in the Kamioka Observatory in Japan \cite{CANDLES}. The main advantage of $^{48}$Ca is that it has the highest Q-value (4.3 MeV) among all the candidate isotopes for 0$\nu\beta\beta$. We are aiming to develop an inorganic scintillator involving $^{48}$Ca with a light yield higher than the 10,000 ph./MeV produced by undoped CaF$_2$, for use in future 0$\nu\beta\beta$ searches.

In this paper, we evaluate the performance of a calcium iodide (CaI$_2$) scintillator involving $^{48}$Ca, with a particular focus on its pulse shape discrimination capability.

\section{Calcium Iodide (CaI$_2$) Scintillator}

CaI$_2$ crystals were discovered by Hofstadter et al. in the 1960s \cite{Hofstadter} and are known to have high light yields, double those of NaI(Tl) scintillators. However, despite their high light yields, these crystals did not come into widespread use due to limitations in early crystal growth and processing techniques.

\begin{figure}[htbp]
 \begin{center}
 \includegraphics[width=5.5cm]{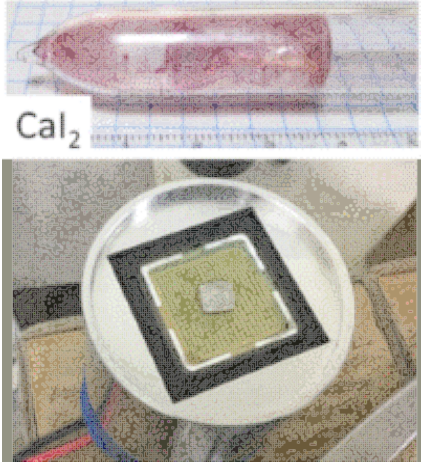}
\caption{Photographs of a CaI$_2$ scintillator crystal grown by the Bridgman--Stockbarger method (top) and the measurement setup (bottom). Here, a CaI$_2$ crystal a few millimeters in size was placed on a photomultiplier tube in a dry room at IMR.}
 \label{fig:picture}
 \end{center}
\end{figure}

Since 2016, the University of Tsukuba and the Institute for Material Research (IMR) at Tohoku University have been jointly developing CaI$_2$ crystals using updated facilities and leading-edge techniques. Initially, it was difficult to vaporize CaI$_2$ due to the similarity of its melting and boiling points. We eventually overcame this problem by sealing off a quartz tube to prevent evaporation, finally succeeding in growing the half-inch CaI$_2$ crystal shown in Figure \ref{fig:picture} by the Bridgman--Stockbarger method \cite{BS}. Here, the pink color seen on the crystal's surface comes from SiCl$_4$, which we used as a reactive-gas atmosphere to eliminate water and oxygen. 

We cut and polished the CaI$_2$ crystal in a dry room (maintained at below 3\% humidity) to create one 5 $\times$ 5 $\times$ 1 mm in size. Then, we measured the crystal's energy response when coupled to photomultiplier tubes (Hamamatsu R7600U-200) with ultra-bialkali photo-cathodes using a $^{137}$Cs gamma-ray source.

\begin{figure}[htbp]
 \begin{center}
 \includegraphics[width=8cm]{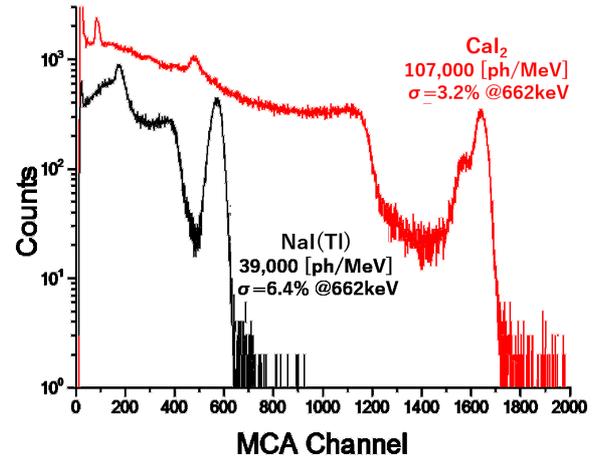}
\caption{Energy spectra for CaI$_2$ and NaI(Tl), produced with 662-keV gamma rays from a $^{137}$Cs source \cite{Kamada}. Here, the photoelectric peak of CaI$_2$ is 2.7 times higher than that of NaI(Tl), and its estimated light yield is 107,000 ph./MeV.}
 \label{fig:spectrum}
 \end{center}
\end{figure}

As Figure \ref{fig:spectrum} shows, using 662-keV gamma rays the photoelectric peak of CaI$_2$ was 2.7 times higher than that of NaI(Tl), and its estimated light yield was 107,000 ph./MeV. We also obtained good resolution, namely 3.2\% @ 662 keV. Table 1 and \cite{Kamada} summarize the measurement results and compare them with those for other scintillators.

\begin{table}[htb]
\begin{center}
    \caption{Summary of the scintillation properties of NaI(Tl) and CaI$_2$ scintillators.}
  \begin{tabular}{ccc}
  	\\ \hline
        & NaI(Tl) & CaI$_2$ \\ \hline \hline
    Light yield   & 39,000 ph./MeV & 107,000 ph./MeV \\ \hline
   Energy resolution   & 6.4 \%@662 keV & 3.2 \%@662 keV \\ \hline
    Decay time  & 230 ns  & 834 ns \\ \hline
     Emission wavelength   & 420 nm & 410 ns \\ \hline
     Density   & 3.67 g/cm$^3$ & 3.97 g/cm$^3$ \\ \hline
  \end{tabular}
\end{center}
\end{table}

\section{Pulse Shape Discrimination Performance of CaI$_2$}

In dark matter search experiments, we need to separate the nuclear recoil signal produced by WIMP scattering from the environmental gamma-ray background. Double beta decay searches use tagged alpha rays produced by U/Th series elements to identify the background from internal radioactive impurities. The detector's particle identification performance is therefore of great importance in astroparticle experiments. Since inorganic scintillator detectors use pulse shape discrimination to separate the signal from the background and improve the signal-to-noise ratio, we now evaluate the pulse shape discrimination performance of our CaI$_2$ crystals.

\begin{figure}[htbp]
 \begin{center}
 \includegraphics[width=8cm]{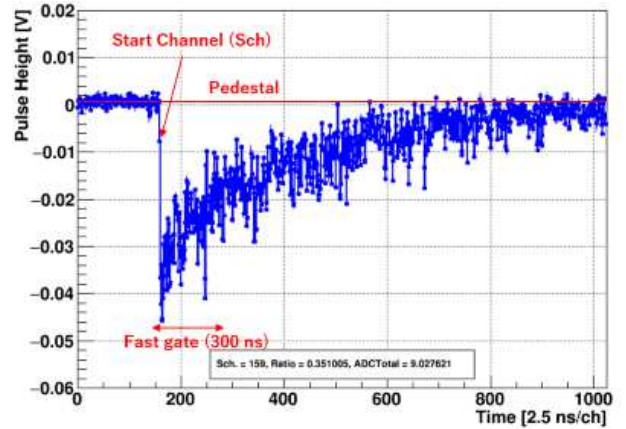}
\caption{Typical CaI$_2$ scintillation light waveform, produced by our setup under $^{137}$Cs gamma-ray excitation.}
 \label{fig:pulse}
 \end{center}
\end{figure}

We took measurements using an $^{241}$Am alpha-ray source and a $^{137}$Cs gamma-ray source and compared the resulting waveforms. The signal from the photomultiplier tube was digitized and acquired at a sampling rate of 400 MHz by a WaveCatcher module \cite{WaveCatcher} connected to a Windows PC. Ten thousand events were recorded by this setup for each source. For each recorded event, we determined the pedestal over the first 100 ch, then took the start channel, i.e., the pulse's rising point, as the first point where the signal continuously exceeded 5 mV from the pedestal for 5 ch. Figure \ref{fig:pulse} shows a typical $^{137}$Cs waveform. 

\begin{figure}[htbp]
 \begin{center}
 \includegraphics[width=8cm]{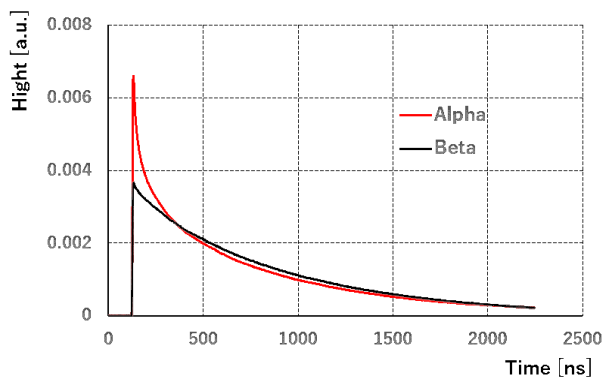}
\caption{Average pulse shapes for alpha (red) and beta (black) ray events. }
 \label{fig:pulse2}
 \end{center}
\end{figure}

\begin{figure}[htbp]
 \begin{center}
 \includegraphics[width=8cm]{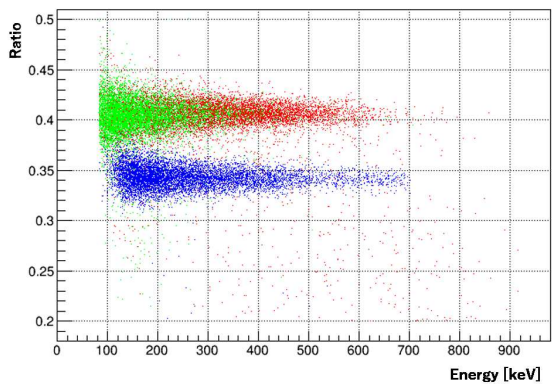}
\caption{Two-dimensional plot of the ratio parameter as a function of energy. Here, the blue, red, and green points represent beta rays, alpha rays, and alpha rays after crystal deliquescence, respectively.}
 \label{fig:ratio}
 \end{center}
\end{figure}

Figure \ref{fig:pulse2} shows the average waveform for ten thousand events. Here, the red line represents alpha-ray events from $^{241}$Am, while the black line represents beta rays from Compton scattering by $^{137}$Cs gamma rays. These curves have been normalized to enclose the same area. The slow components are essentially the same for both the alpha and beta rays, but there is a substantial difference over the first 300 ns or so. For our pulse shape discrimination analysis, we therefore defined a ratio parameter based on the ratio of the area for the first 300 ns to the total area. Figure \ref{fig:ratio} shows the ratio parameter as a function of energy, based on data for 662-keV gamma rays from $^{137}$Cs, for beta (blue) and alpha (red) rays, respectively. This approach enabled us to achieve very good separation, at least above 100 keV.

CaI$_2$ has strong hygroscopic properties and reacts with water in the air to form CaI$_2\cdot$(H$_2$O)$_n$. Even in a dry room, the surface of a CaI$_2$ crystal deliquesces and becomes white within an hour. Because alpha rays are stopped by the surface, we believed that differences between the alpha and beta rays would appear over time due to the impact of surface deliquescence. In order to investigate this hypothesis, we took measurements for a deliquesced CaI$_2$ crystal that had been left in a dry room for more than an hour; the results are shown in Figure \ref{fig:ratio} in green. These results indicate that deliquescence does not change the pulse shape, but it does decrease light yield. Thus, the strong pulse shape discrimination capability of CaI$_2$ is unaffected by deliquescence.

\section{Conclusion and Prospects}
Inorganic scintillators are being used for dark matter and double beta decay experiments. We are thus developing CaI$_2$ crystals that contain $^{48}$Ca from nuclei that have undergone double beta decay and can generate high luminescence yields. We prepared half-inch CaI$_2$ crystals by the Bridgman--Stockbarger method at IMR, then sliced off samples a few millimeters square to measure their scintillation characteristics. The amount of light they emitted was 107,000 ph./MeV, 2.7 times more than NaI(Tl) and 10 times more than undoped CaF$_2$. In addition, the light was emitted at a wavelength of 410 nm, within the photomultiplier's sensitivity range. When we measured the crystal's waveform discrimination characteristics using $^{241}$Am alpha rays and $^{137}$Cs gamma rays, we found substantial differences in the waveforms over the first 300 ns. Thus, even a simple ratio analysis achieved very good separation, meaning that CaI$_2$ shows great potential for pulse shape discrimination.

At present, it is difficult to grow and process large crystals due to the rapid deliquescence and frequent cleavage of CaI$_2$. Research is underway to reduce cleavage by changing the crystal growth parameters and crystal composition. In future work, we will also develop an analysis method that can achieve better pulse shape discrimination and use it to create a detector.

\section*{Acknowledgment} \label{acknowledgment}

This work was performed as part of the Inter-University Cooperative Research Program of the Institute for Materials Research, Tohoku University (Proposal No. 18K0061 and 19K0089). It was supported by JSPS KAKENHI Grant-in-Aid for Scientific Research (B) 18H01222 and Grant-in-Aid for Young Scientists (B) 16K17700. The authors also appreciate the support of the University of Tsukuba's Basic Research Support Program Type S.

\section*{References}

\bibliography{mybibfile}

\end{document}